\documentclass[12pt]{article}

\usepackage[letterpaper, margin=1in]{geometry}

\usepackage[colorlinks]{hyperref}
\usepackage{makeidx}
\usepackage{framed}
\usepackage{fancyvrb}
\usepackage{color}
\usepackage{hyperref}
\hypersetup{colorlinks = true, allcolors = blue}
\usepackage{float}
\usepackage{flafter}
\usepackage{tabularx}
\usepackage{booktabs}
\usepackage{longtable}
\usepackage{multirow}
\usepackage{graphicx}
\usepackage{afterpage}

\usepackage{datetime}

\yyyymmdddate

  \usepackage{setspace}
  \AtBeginEnvironment{tabular}{\singlespacing}
  \AtBeginEnvironment{table}{\singlespacing}
  \AtBeginEnvironment{longtable}{\singlespacing}
  \AtBeginEnvironment{verbatim}{\singlespacing}
  \AtBeginEnvironment{Shaded}{\singlespacing}

\NewDocumentCommand\citeproctext{}{}

\makeatletter
 \let\@cite@ofmt\@firstofone
 \def\@biblabel#1{}
 \def\@cite#1#2{{#1\if@tempswa , #2\fi}}
\makeatother
\newlength{\cslhangindent}
\newlength{\csllabelwidth}
\newenvironment{CSLReferences}[2] 
 {\begin{list}{}{%
  \setlength{\itemindent}{0pt}
  \setlength{\leftmargin}{0pt}
  \setlength{\parsep}{0pt}
  \ifodd #1
   \setlength{\leftmargin}{\cslhangindent}
   \setlength{\itemindent}{-1\cslhangindent}
  \fi
  \setlength{\itemsep}{#2\baselineskip}}}
 {\end{list}}
\usepackage{calc}

  \usepackage{amsmath}
  \usepackage{amssymb}

\usepackage{xcolor}
\definecolor{dark}{HTML}{2c2e35}
\color{dark}

\definecolor{myblue}{HTML}{1e3765}
\hypersetup{colorlinks,linkcolor=myblue,urlcolor=myblue}

\makeindex

\providecommand{\tightlist}{\setlength{\itemsep}{0pt}\setlength{\parskip}{0pt}}

  \usepackage{color}
  \usepackage{fancyvrb}

  \DefineVerbatimEnvironment{Highlighting}{Verbatim}{commandchars=\\\{\}}
  \usepackage{framed}
  \definecolor{shadecolor}{RGB}{241,243,245}
  \newenvironment{Shaded}{\begin{snugshade}}{\end{snugshade}}

  \newcommand{\AttributeTok}[1]{\textcolor[rgb]{0.40,0.45,0.13}{#1}}

  \newcommand{\CommentTok}[1]{\textcolor[rgb]{0.37,0.37,0.37}{#1}}

  \newcommand{\ControlFlowTok}[1]{\textcolor[rgb]{0.00,0.23,0.31}{\textbf{#1}}}
  \newcommand{\DataTypeTok}[1]{\textcolor[rgb]{0.68,0.00,0.00}{#1}}
  \newcommand{\DecValTok}[1]{\textcolor[rgb]{0.68,0.00,0.00}{#1}}
  
  \newcommand{\ErrorTok}[1]{\textcolor[rgb]{0.68,0.00,0.00}{#1}}
  
  \newcommand{\FloatTok}[1]{\textcolor[rgb]{0.68,0.00,0.00}{#1}}
  \newcommand{\FunctionTok}[1]{\textcolor[rgb]{0.28,0.35,0.67}{#1}}
  \newcommand{\ImportTok}[1]{\textcolor[rgb]{0.00,0.46,0.62}{#1}}
  
  \newcommand{\KeywordTok}[1]{\textcolor[rgb]{0.00,0.23,0.31}{\textbf{#1}}}
  \newcommand{\NormalTok}[1]{\textcolor[rgb]{0.00,0.23,0.31}{#1}}
  \newcommand{\OperatorTok}[1]{\textcolor[rgb]{0.37,0.37,0.37}{#1}}
  \newcommand{\OtherTok}[1]{\textcolor[rgb]{0.00,0.23,0.31}{#1}}
  \newcommand{\PreprocessorTok}[1]{\textcolor[rgb]{0.68,0.00,0.00}{#1}}
  
  \newcommand{\SpecialCharTok}[1]{\textcolor[rgb]{0.37,0.37,0.37}{#1}}
  
  \newcommand{\StringTok}[1]{\textcolor[rgb]{0.13,0.47,0.30}{#1}}
  \newcommand{\VariableTok}[1]{\textcolor[rgb]{0.07,0.07,0.07}{#1}}

\author{
  Mauricio Vargas Sepúlveda (ORCID 0000-0003-1017-7574)\\Department of
Political Science, University of Toronto\\Munk School of Global Affairs
and Public Policy, University of Toronto\\
  \smallskip\\
  Jonathan Schneider Malamud\\Department of Electrical and Computer
Engineering, University of Toronto\\
  \smallskip\\
  Corresponding author: m.sepulveda@mail.utoronto.ca
}
\title{cpp11armadillo: An R Package to Use the Armadillo C++ Library}
\date{Last updated: \today\ \currenttime}

\begin{document}

\maketitle

\thispagestyle{empty}
\tableofcontents
\setcounter{page}{0}
\clearpage

\afterpage{\setlength\parskip{10pt}}

\section{Abstract}\label{abstract}

This article introduces `cpp11armadillo', a new R package that
integrates the powerful Armadillo C++ library for linear algebra into
the R programming environment. Targeted primarily at social scientists
and other non-programmers, this article explains the computational
benefits of moving code to C++ in terms of speed and syntax. We provide
a comprehensive overview of Armadillo's capabilities, highlighting its
user-friendly syntax akin to MATLAB and its efficiency for
computationally intensive tasks. The `cpp11armadillo' package simplifies
a part of the process of using C++ within R by offering additional ease
of integration for those who require high-performance linear algebra
operations in their R workflows. This work aims to bridge the gap
between computational efficiency and accessibility, making advanced
linear algebra operations more approachable for R users without
extensive programming backgrounds.

\section{Introduction}\label{introduction}

\texttt{R} is widely used by non-programmers (Wickham et al. 2019), and
this article aims to introduce computational concepts in a non-technical
yet formal manner for social scientists. Our goal is to explain when and
why moving code to C++ is beneficial in terms of speed or syntax and how
to do it using
\href{https://pacha.dev/cpp11armadillo}{\texttt{cpp11armadillo}}, our
novel Armadillo and \texttt{R} integration for linear algebra.

\href{https://arma.sourceforge.net/}{Armadillo} is a C++ library
designed for linear algebra, emphasizing a balance between performance
and ease of use. C++ is highly efficient for computationally intensive
tasks but lacks built-in data structures and functions for linear
algebra operations. Armadillo fills this gap by providing an intuitive
syntax similar to MATLAB (Sanderson and Curtin 2016).

\href{https://cran.r-project.org/package=RcppArmadillo}{\texttt{rcpparmadillo}},
introduced in 2010, integrates Armadillo with \texttt{R} through the
Rcpp package, enabling the use of C++ for performance-critical parts of
\texttt{R} code (Eddelbuettel and Sanderson 2014).
\texttt{rcpparmadillo} is a widely successful project, and at the time
of writing this article, there are 755 packages on CRAN that depend on
it (Lee 2024).

\texttt{cpp11armadillo} is an independent project that aims to simplify
the integration of \texttt{R} and C++ by using \texttt{cpp11}, an
\texttt{R} package that eases using C++ functions from \texttt{R}, and
it is aligned with the tidyverse philosophy of simplicity and
user-centric design (Wickham et al. 2019; Vaughan, Hester, and François
2023). Is it useful in cases where vectorization (e.g., applying an
operation to a vector or matrix as a whole rather than looping over each
element) is not possible or challenging, and it can help to solve some
bottlenecks as it simplifies the task of rewriting \texttt{R} code that
involves linear algebra as C++ code. Furthermore, it can be orders of
magnitude faster, computing operations in parallel, which is especially
useful for large objects. When vectorization is possible, using
\texttt{R}`s built-in functions is more efficient than writing loops in
\texttt{R}, and the time of writing the same in C++ justifies to
continue to use vectorized operations in \texttt{R}. 'cpp11armadillo' is
useful in cases where vectorization (e.g., applying an operation to a
vector or matrix as a whole rather than looping over each element) is
not possible or challenging, and it can help to solve some bottlenecks
as it simplifies the task of rewriting R code that involves linear
algebra as C++ code.

For cases where vectorization is applicable, Burns (2011) provides a
good introduction. \texttt{cpp11armadillo} is relevant in a project
where the same operation is repeated many times and, at the same time,
the computation time saved by using C++ is greater than the time spent
writing and fixing C++ code. We followed four design principles when
developing \texttt{cpp11armadillo}:

\begin{enumerate}
\def\labelenumi{\arabic{enumi}.}
\tightlist
\item
  \textbf{Column oriented}: The code and documentation uses column
  vectors, following Hansen (2022).
\item
  \textbf{Package oriented}: It is designed to be used in an \texttt{R}
  package, which is the recommended way to organize code in medium and
  large scale projects.
\item
  \textbf{Header only}: No separate actions are required; it only
  requires to include \texttt{cpp11armadillo} as a dependency.
\item
  \textbf{Vendoring capable}: It provides a dedicated function to copy
  the entire codebase into \texttt{R} packages, providing the option to
  make it a one-time dependency. This feature allows to run code in
  restricted environments (e.g., where installing packages from CRAN or
  GitHub is blocked by a firewall or available for administrators only).
  \# Interpreted and compiled languages
\end{enumerate}

\texttt{R} is an interpreted language, meaning that the code is executed
line by line when you run a part or the totality of a script. One
advantage of interpreted languages is that they are easier to debug
because the code can be run by parts to isolate errors. Another
advantage is that, assuming that all dependencies are solved, the code
can be run in any computer without additional configurations. Other
interpreted languages are Python, MATLAB, and Wolfram.

C++ is a compiled language, meaning that to run the code, it must be
converted to an executable file containing instructions that the
processor can understand. This allows the compiler to optimize the code
for the specific hardware it is running on, making it faster than
interpreted languages. The main disadvantage of compiled languages is
that they are harder to debug because it is not possible to run the code
by parts as when running \texttt{R} code blocks on-the-fly. C++ code
requires a compiler (e.g.~gcc or clang) to produce an executable file,
which is a software separate from the editor (e.g. RStudio or VS Code)
that translates the code to machine code, and it is not possible to use
an executable produced on Windows with UNIX and vice versa. Other
compiled languages are C, FORTRAN, and Java.

\texttt{R} internals consist in functions written in C and FORTRAN that
the end user has ready-made to run scripts. These functions, while
available in \texttt{R} source code, are usually not of interest for the
end user as these already have a proven stability, even in corner cases,
and those are written with memory and speed efficiency at the expense of
syntax and flexibility.

C inherent steep learning curve motivated C++ creation, and C++ is a
superset of C with additional features that make it easier to use and
that provides flexibility. C++ is a high-level language that can be used
for a wide range of purposes, including parts of operating systems
(e.g., Windows), internet browsers (e.g., Firefox) and streaming
platforms (e.g., YouTube), and it is particularly useful for
computationally intensive tasks. Transitioning from \texttt{R} to C++
involves adapting to several differences in conventions and flexibility
regarding data types and operations.

\section{Linear algebra libraries}\label{linear-algebra-libraries}

Linear algebra libraries are essential for scientific computing, and
they provide functions for matrix operations, such as matrix
multiplication, inversion, and decomposition. These libraries are
written in C or C++ and are optimized for speed without compromising
stability. Some of these libraries are the Linear Algebra Package
(LAPACK) and the Basic Linear Algebra Subprograms (BLAS). As C and C++,
these libraries have a long history and time-tested correctness, and the
first LAPACK release was in 1992 and the first BLAS release was in 1979.

When you run \texttt{sessionInfo()} in \texttt{R}, it shows lines
similar to:

\begin{Shaded}
\begin{Highlighting}[]
\NormalTok{Matrix products}\SpecialCharTok{:}\NormalTok{ default}
\NormalTok{BLAS}\SpecialCharTok{:}   \ErrorTok{/}\NormalTok{usr}\SpecialCharTok{/}\NormalTok{lib}\SpecialCharTok{/}\NormalTok{x86\_64}\SpecialCharTok{{-}}\NormalTok{linux}\SpecialCharTok{{-}}\NormalTok{gnu}\SpecialCharTok{/}\NormalTok{blas}\SpecialCharTok{/}\NormalTok{libblas.so.}\DecValTok{3}\NormalTok{.}\FloatTok{10.0} 
\NormalTok{LAPACK}\SpecialCharTok{:} \ErrorTok{/}\NormalTok{usr}\SpecialCharTok{/}\NormalTok{lib}\SpecialCharTok{/}\NormalTok{x86\_64}\SpecialCharTok{{-}}\NormalTok{linux}\SpecialCharTok{{-}}\NormalTok{gnu}\SpecialCharTok{/}\NormalTok{lapack}\SpecialCharTok{/}\NormalTok{liblapack.so.}\DecValTok{3}\NormalTok{.}\FloatTok{10.0}
\end{Highlighting}
\end{Shaded}

Or similar to:

\begin{Shaded}
\begin{Highlighting}[]
\NormalTok{Matrix products}\SpecialCharTok{:}\NormalTok{ default}
\NormalTok{BLAS}\SpecialCharTok{:}   \ErrorTok{/}\NormalTok{usr}\SpecialCharTok{/}\NormalTok{lib}\SpecialCharTok{/}\NormalTok{x86\_64}\SpecialCharTok{{-}}\NormalTok{linux}\SpecialCharTok{{-}}\NormalTok{gnu}\SpecialCharTok{/}\NormalTok{openblas}\SpecialCharTok{{-}}\NormalTok{pthread}\SpecialCharTok{/}\NormalTok{libblas.so}\FloatTok{.3} 
\NormalTok{LAPACK}\SpecialCharTok{:} \ErrorTok{/}\NormalTok{usr}\SpecialCharTok{/}\NormalTok{lib}\SpecialCharTok{/}\NormalTok{x86\_64}\SpecialCharTok{{-}}\NormalTok{linux}\SpecialCharTok{{-}}\NormalTok{gnu}\SpecialCharTok{/}\NormalTok{openblas}\SpecialCharTok{{-}}\NormalTok{pthread}\SpecialCharTok{/}\NormalTok{libopenblasp}\SpecialCharTok{{-}}\NormalTok{r0.}\DecValTok{3}\NormalTok{.}\FloatTok{20.}\NormalTok{so;}
\NormalTok{LAPACK version }\DecValTok{3}\NormalTok{.}\FloatTok{10.0}
\end{Highlighting}
\end{Shaded}

This information reveals that \texttt{R} is using the BLAS and LAPACK
libraries for linear algebra operations. BLAS and LAPACK are used
internally when executing functions such as \texttt{lm()},
\texttt{solve()} or \texttt{\%*\%}.

Armadillo also calls BLAS and LAPACK for linear algebra operations, and
this can be verified in its source code that contains the lines:

\begin{Shaded}
\begin{Highlighting}[]
\PreprocessorTok{\#include }\ImportTok{"armadillo/def\_blas.hpp"}
\PreprocessorTok{\#include }\ImportTok{"armadillo/def\_lapack.hpp"}
\end{Highlighting}
\end{Shaded}

While it is possible to use BLAS or LAPACK directly in C++ code,
Armadillo provides efficient routines that largely simplify the syntax
and the time involved to write useable code, and it combines operations
to reduce intermediate steps and temporary objects when possible.
Armadillo is not just about speed, in some cases it is also about
feasibility, as it allows to use the available memory more efficiently
and to run tasks that would be impossible with the available memory if
done in a naive way that involves creating temporary objects and copies.

Just as an elemental example, without additional details, the following
code computes the dot product of two vectors using BLAS:

\begin{Shaded}
\begin{Highlighting}[]
\CommentTok{// 2x2 matrix written as a long array}
\DataTypeTok{double}\NormalTok{ a}\OperatorTok{[}\DecValTok{4}\OperatorTok{]} \OperatorTok{=} \OperatorTok{\{}\FloatTok{1.0}\OperatorTok{,} \FloatTok{2.0}\OperatorTok{,} \FloatTok{3.0}\OperatorTok{,} \FloatTok{4.0}\OperatorTok{\};}
    
\CommentTok{// matrix to write the transposed result to}
\DataTypeTok{double}\NormalTok{ at}\OperatorTok{[}\DecValTok{4}\OperatorTok{];}

\CommentTok{// dimensions}
\DataTypeTok{int}\NormalTok{ r }\OperatorTok{=} \DecValTok{2}\OperatorTok{,}\NormalTok{ c }\OperatorTok{=} \DecValTok{2}\OperatorTok{;}

\CommentTok{// transpose}
\NormalTok{cblas\_dgeam}\OperatorTok{(}\NormalTok{CblasRowMajor}\OperatorTok{,}\NormalTok{ CblasTrans}\OperatorTok{,}\NormalTok{ CblasTrans}\OperatorTok{,}\NormalTok{ rows}\OperatorTok{,}\NormalTok{ cols}\OperatorTok{,} 
  \FloatTok{1.0}\OperatorTok{,}\NormalTok{ matrix}\OperatorTok{,}\NormalTok{ cols}\OperatorTok{,} \FloatTok{0.0}\OperatorTok{,} \KeywordTok{nullptr}\OperatorTok{,}\NormalTok{ cols}\OperatorTok{,}\NormalTok{ transposed\_matrix}\OperatorTok{,}\NormalTok{ rows}\OperatorTok{);}
\end{Highlighting}
\end{Shaded}

The same result can be obtained with \texttt{R}:

\begin{Shaded}
\begin{Highlighting}[]
\NormalTok{A }\OtherTok{\textless{}{-}} \FunctionTok{matrix}\NormalTok{(}\FunctionTok{c}\NormalTok{(}\DecValTok{1}\NormalTok{, }\DecValTok{2}\NormalTok{, }\DecValTok{3}\NormalTok{, }\DecValTok{4}\NormalTok{), }\AttributeTok{nrow =} \DecValTok{2}\NormalTok{)}
\NormalTok{At }\OtherTok{\textless{}{-}} \FunctionTok{t}\NormalTok{(A)}
\end{Highlighting}
\end{Shaded}

The same result can be obtained with Armadillo:

\begin{Shaded}
\begin{Highlighting}[]
\NormalTok{Mat}\OperatorTok{\textless{}}\DataTypeTok{double}\OperatorTok{\textgreater{}}\NormalTok{ A }\OperatorTok{=} \OperatorTok{\{\{}\FloatTok{1.0}\OperatorTok{,} \FloatTok{2.0}\OperatorTok{\},} \OperatorTok{\{}\FloatTok{3.0}\OperatorTok{,} \FloatTok{4.0}\OperatorTok{\}\};}
\NormalTok{Mat}\OperatorTok{\textless{}}\DataTypeTok{double}\OperatorTok{\textgreater{}}\NormalTok{ At }\OperatorTok{=}\NormalTok{ A}\OperatorTok{.}\NormalTok{t}\OperatorTok{();}
\end{Highlighting}
\end{Shaded}

Writing code directly in BLAS or LAPACK can be challenging, and it is
very challenging to beat BLAS or LAPACK performance. This is why
\texttt{R} and Armadillo use them for the internal computation and
provide functions with a simplified syntax for the end user. The
interested readers can explore Zhang and Kroeker (2024) to read about
OpenBLAS, an even faster version of BLAS.

\section{R vectorization and loops}\label{r-vectorization-and-loops}

Some of \texttt{R}'s vectorized functions include \texttt{sum()},
\texttt{mean()}, \texttt{apply()} and its variants, and \texttt{map()}
and its variants in the \texttt{purrr} package (`R' Core Team 2024;
Wickham et al. 2019).

Instead of using the \texttt{mean()} function, the mean of a vector can
be computed with a loop in \texttt{R}:

\begin{Shaded}
\begin{Highlighting}[]
\NormalTok{x }\OtherTok{\textless{}{-}} \FunctionTok{c}\NormalTok{(}\DecValTok{1}\NormalTok{, }\DecValTok{2}\NormalTok{, }\DecValTok{3}\NormalTok{, }\DecValTok{4}\NormalTok{, }\DecValTok{5}\NormalTok{)}

\NormalTok{numerator }\OtherTok{\textless{}{-}} \DecValTok{0}
\NormalTok{denominator }\OtherTok{\textless{}{-}} \FunctionTok{length}\NormalTok{(x)}

\ControlFlowTok{for}\NormalTok{ (i }\ControlFlowTok{in} \DecValTok{1}\SpecialCharTok{:}\NormalTok{denominator) \{}
\NormalTok{  y }\OtherTok{\textless{}{-}}\NormalTok{ y }\SpecialCharTok{+}\NormalTok{ x[i]}
\NormalTok{\}}

\NormalTok{numerator }\SpecialCharTok{/}\NormalTok{ denominator}
\end{Highlighting}
\end{Shaded}

The previous loop is inefficient because it involves writing more code
and it is slower because it goes through each element one by one. In
\texttt{R} or any other interpreted language (e.g.~Python), loops are
slower than vectorized operations.

An example of efficient vectorization is the \texttt{pmax()} function to
obtain the element-wise maximum for vectors or matrices. The same can be
done for two input matrices with a loop in \texttt{R} that has the
advantage of being explicit in terms of the operations performed, but it
is slower than the vectorized function:

\begin{Shaded}
\begin{Highlighting}[]
\NormalTok{A }\OtherTok{\textless{}{-}} \FunctionTok{matrix}\NormalTok{(}\FunctionTok{c}\NormalTok{(}\DecValTok{1}\NormalTok{, }\DecValTok{2}\NormalTok{, }\DecValTok{3}\NormalTok{, }\DecValTok{4}\NormalTok{), }\AttributeTok{nrow =} \DecValTok{2}\NormalTok{)}
\NormalTok{B }\OtherTok{\textless{}{-}} \FunctionTok{matrix}\NormalTok{(}\FunctionTok{c}\NormalTok{(}\DecValTok{4}\NormalTok{, }\DecValTok{3}\NormalTok{, }\DecValTok{2}\NormalTok{, }\DecValTok{1}\NormalTok{), }\AttributeTok{nrow =} \DecValTok{2}\NormalTok{)}

\NormalTok{C }\OtherTok{\textless{}{-}} \FunctionTok{matrix}\NormalTok{(}\DecValTok{0}\NormalTok{, }\AttributeTok{nrow =} \DecValTok{2}\NormalTok{, }\AttributeTok{ncol =} \DecValTok{2}\NormalTok{)}

\ControlFlowTok{for}\NormalTok{ (i }\ControlFlowTok{in} \DecValTok{1}\SpecialCharTok{:}\DecValTok{2}\NormalTok{) \{}
  \ControlFlowTok{for}\NormalTok{ (j }\ControlFlowTok{in} \DecValTok{1}\SpecialCharTok{:}\DecValTok{2}\NormalTok{) \{}
\NormalTok{    C[i, j] }\OtherTok{\textless{}{-}} \FunctionTok{max}\NormalTok{(A[i, j], B[i, j])}
\NormalTok{  \}}
\NormalTok{\}}
\end{Highlighting}
\end{Shaded}

The problem with this loop is that it is particularly slow, for two
\(1000\times 1000\) matrices filled random values created with
\texttt{rnorm()}, it takes takes 422 miliseconds to run, while the
\texttt{pmax()} function takes 7 miliseconds. In other words, the
implemented loop is twenty times slower.

Loops should be used in computations where the one step depends on the
previous steps. One example of this is the Gram-Schmidt method to obtain
an orthogonal matrix from a square matrix \(X\), in which case the
\(N^{th}\) vector depends on the previous \(1,2,\ldots,N-1\) vectors,
and it consists of the following algorithm adapted from Strang (1988):

\begin{itemize}
\tightlist
\item
  Step 1: Construct a matrix \(X\) of dimension \(M \times N\) with the
  vectors to be orthonormalized as column vectors.
\item
  Step 2: Construct a matrix \(U\) of the same dimension as \(X\) filled
  with zeroes to store the orthonormal basis later.
\item
  Step 3: Replace the first column of \(U\) with the vector
  \(u_1 = x_1 / \| x_1 \|\), where \(x_1\) is the first column of \(X\)
  and \(\| x_1 \|\) is the euclidean norm that is the square root of the
  sum of the squared \(m\) coordinates \(x_{1i}\) given by
  \(\sqrt{\sum_{i=1}^m x_{1i}^2} = \sqrt{x_1^t x_1}\).
\item
  Step 4: For the remaining \(M-1\) vectors \(x_{j>1}\), calculate the
  projection of the vector \(x_j\) onto the vector \(u_j\) and subtracts
  it from \(x_j\), this is
  \(x_j = x_j - \sum_{i=1}^{j-1} (u_i^t x_j / u_i^t u_i) u_i\).
\item
  Step 5: Normalize each \(x_{j>1}\) to unit length as
  \(x_j = x_j / \|x_j\|\) and replace it in the remaining columns of
  \(U\).
\end{itemize}

In \texttt{R} this can be written as:

\begin{Shaded}
\begin{Highlighting}[]
\NormalTok{X }\OtherTok{\textless{}{-}} \FunctionTok{matrix}\NormalTok{(}\FunctionTok{c}\NormalTok{(}\DecValTok{3}\NormalTok{,}\DecValTok{4}\NormalTok{,}\DecValTok{4}\NormalTok{,}\DecValTok{4}\NormalTok{), }\AttributeTok{nrow =} \DecValTok{2}\NormalTok{)}
\NormalTok{U }\OtherTok{\textless{}{-}} \FunctionTok{matrix}\NormalTok{(}\DecValTok{0}\NormalTok{, }\AttributeTok{nrow =} \DecValTok{2}\NormalTok{, }\AttributeTok{ncol =} \DecValTok{2}\NormalTok{)}

\NormalTok{N }\OtherTok{\textless{}{-}} \FunctionTok{ncol}\NormalTok{(X)}

\NormalTok{U[, }\DecValTok{1}\NormalTok{] }\OtherTok{\textless{}{-}}\NormalTok{ X[, }\DecValTok{1}\NormalTok{] }\SpecialCharTok{/} \FunctionTok{sqrt}\NormalTok{(}\FunctionTok{sum}\NormalTok{(X[, }\DecValTok{1}\NormalTok{]}\SpecialCharTok{\^{}}\DecValTok{2}\NormalTok{))}

\ControlFlowTok{for}\NormalTok{ (j }\ControlFlowTok{in} \DecValTok{2}\SpecialCharTok{:}\NormalTok{N) \{}
\NormalTok{  v }\OtherTok{\textless{}{-}}\NormalTok{ X[, j]}
  \ControlFlowTok{for}\NormalTok{ (i }\ControlFlowTok{in} \DecValTok{1}\SpecialCharTok{:}\NormalTok{(j }\SpecialCharTok{{-}} \DecValTok{1}\NormalTok{)) \{}
\NormalTok{    u }\OtherTok{\textless{}{-}}\NormalTok{ U[, i]}
\NormalTok{    v }\OtherTok{\textless{}{-}}\NormalTok{ v }\SpecialCharTok{{-}}\NormalTok{ (}\FunctionTok{crossprod}\NormalTok{(u, v) }\SpecialCharTok{/} \FunctionTok{crossprod}\NormalTok{(u, u)) }\SpecialCharTok{*}\NormalTok{ u}
\NormalTok{  \}}
\NormalTok{  U[, j] }\OtherTok{\textless{}{-}}\NormalTok{ v }\SpecialCharTok{/} \FunctionTok{sqrt}\NormalTok{(}\FunctionTok{sum}\NormalTok{(v}\SpecialCharTok{\^{}}\DecValTok{2}\NormalTok{))}
\NormalTok{\}}
\end{Highlighting}
\end{Shaded}

This result is correct according to Wolfram Alpha, that returns the
column vectors \(c_1 = (3/5,4/5)\) and \(c_2 = (4/5,-3/5)\).

\section{Common pitfalls when transitioning from R to
C++}\label{common-pitfalls-when-transitioning-from-r-to-c}

\subsection{Syntax and defaults}\label{syntax-and-defaults}

Semicolons are mandatory in C++. C++ defaults to int for numbers, while
\texttt{R} defaults to double. In C++, you must declare the variable
type.

The following \texttt{R} code treats \(x\) as a double (e.g.~a decimal
number) unless otherwise specified:

\begin{Shaded}
\begin{Highlighting}[]
\CommentTok{\# double}
\NormalTok{x }\OtherTok{\textless{}{-}} \DecValTok{200}
\ControlFlowTok{function}\NormalTok{(x) \{}
\NormalTok{  x }\SpecialCharTok{+} \DecValTok{100}
\NormalTok{\}}

\CommentTok{\# integer}
\NormalTok{x }\OtherTok{\textless{}{-}} \DecValTok{200}\NormalTok{L}
\ControlFlowTok{function}\NormalTok{(x) \{}
\NormalTok{  x }\SpecialCharTok{+} \DecValTok{100}\NormalTok{L }\CommentTok{\# L = integer}
\NormalTok{\}}
\end{Highlighting}
\end{Shaded}

C++ needs to declare the data type of the variable even if the number
does not have a decimal point:

\begin{Shaded}
\begin{Highlighting}[]
\CommentTok{// integer}
\DataTypeTok{int}\NormalTok{ x }\OperatorTok{=} \DecValTok{200}\OperatorTok{;}
\DataTypeTok{double}\NormalTok{ function}\OperatorTok{(}\DataTypeTok{double}\NormalTok{ y}\OperatorTok{)} \OperatorTok{\{}
  \ControlFlowTok{return}\NormalTok{ y }\OperatorTok{+} \DecValTok{100}\OperatorTok{;}
\OperatorTok{\}}

\CommentTok{// double}
\DataTypeTok{double}\NormalTok{ x }\OperatorTok{=} \FloatTok{200.0}\OperatorTok{;} \CommentTok{// x = 200 also works}
\DataTypeTok{double}\NormalTok{ function}\OperatorTok{(}\DataTypeTok{double}\NormalTok{ y}\OperatorTok{)} \OperatorTok{\{}
  \ControlFlowTok{return}\NormalTok{ y }\OperatorTok{+} \FloatTok{100.0}\OperatorTok{;} \CommentTok{// y + 100 also works}
\OperatorTok{\}}
\end{Highlighting}
\end{Shaded}

In \texttt{R}, variable names can be recycled in any function without
issues. In C++ the example, the function has an argument \texttt{y}
instead of \texttt{x} because \texttt{x} was previously declared in the
global scope, and the code would not compile if the function had an
argument \texttt{x}.

\subsection{Lack of a terminal Shell}\label{lack-of-a-terminal-shell}

C++ lacks a dedicated terminal shell and cannot be used as a scientific
calculator like \texttt{R}. C++ code must be compiled and executed. An
analogy for this is that \texttt{R} is like ready-made Eggo waffles,
while C++ is like making waffles from scratch by mixing the ingredients.

\subsection{Data types}\label{data-types}

C++ requires explicit library inclusion for strings, vectors, matrices,
lists, and data frames, which are not natively available. \texttt{R} has
built-in data structures for these types. \texttt{cpp11} provides
wrappers for these data structures, which facilitate \texttt{R} and C++
integration. This is similar to \texttt{R}'s tibble, a data structure
not natively available in base \texttt{R} but that is provided by the
\texttt{tibble} package, and that enhances the data frame structure
(Müller and Wickham 2023).

In addition to \texttt{cpp11} vectors and matrices, Armadillo provides
its own data structures for linear algebra operations. Armadillo data
structures are more flexible and allow for a highly readable and concise
syntax, this is why \texttt{cpp11armadillo} exists, because \texttt{R}
cannot directly use Armadillo data structures unless there is a package
that translates them to \texttt{R} data structures. This is similar to
\texttt{R}'s SQL integration, where the \texttt{rpostgres} package
contains C++ code capable of translating a SQL query to a tibble
(Wickham, Ooms, and Müller 2023).

\subsection{Operations and indexing}\label{operations-and-indexing}

C++ has useful operators that do not exist in \texttt{R} (e.g., ++, +=,
and *=). C++ is zero-indexed, whereas \texttt{R} is one-indexed.

The following \texttt{R} code sums the numbers in the sequence 5 7 4 4 2

\begin{verbatim}
x <- c(5, 7, 4, 4, 2)

for (i in 1:5) {
  x[i] <- x[i] + i
}
\end{verbatim}

An equivalent C++ code is:

\begin{Shaded}
\begin{Highlighting}[]
\DataTypeTok{int}\NormalTok{ x}\OperatorTok{[}\DecValTok{5}\OperatorTok{]} \OperatorTok{=} \OperatorTok{\{}\DecValTok{5}\OperatorTok{,} \DecValTok{7}\OperatorTok{,} \DecValTok{4}\OperatorTok{,} \DecValTok{4}\OperatorTok{,} \DecValTok{2}\OperatorTok{\};}

\ControlFlowTok{for} \OperatorTok{(}\DataTypeTok{int}\NormalTok{ i }\OperatorTok{=} \DecValTok{0}\OperatorTok{;}\NormalTok{ i }\OperatorTok{\textless{}} \DecValTok{5}\OperatorTok{;} \OperatorTok{++}\NormalTok{i}\OperatorTok{)} \OperatorTok{\{}
\NormalTok{  x}\OperatorTok{[}\NormalTok{i}\OperatorTok{]} \OperatorTok{=}\NormalTok{ x}\OperatorTok{[}\NormalTok{i}\OperatorTok{]} \OperatorTok{+} \OperatorTok{(}\NormalTok{i }\OperatorTok{+} \DecValTok{1}\OperatorTok{);}
\OperatorTok{\}}
\end{Highlighting}
\end{Shaded}

\section{Computational complexity}\label{computational-complexity}

Computational complexity refers to the number of steps required to solve
a problem. It is expressed in terms of the size of the input data,
\(n\), and the number of operations required to solve the problem.

The same algorithm implemented in different programming languages will
retain its computational complexity. Rewriting an \texttt{R} code in C++
may reduce the chronological time to run a function, but the complexity
will remain the same. The only possibility to reduce the computational
complexity is to write an equivalent algorithm that implement different
steps to do the same, which is why reduced forms are important in the
field of Econometrics and others.

C++ is faster for loops because the time (e.g., seconds) it takes for
each iteration of the loop is usually lower than in \texttt{R}, but for
equivalent loops in C++ and \texttt{R} the total number of operations is
the same. One of the notations, the big-O notation is expressed as
\(O(f(n))\), where \(f(n)\) is a function that describes the upper bound
of the number of operations required to solve the problem.

For the previous loop to compute the mean of a vector of \(n\)
coordinates (or elements), the computational complexity is \(O(n)\)
because there are \(n\) elements to sum to create the numerator plus one
division by the denominator given by the number of elements, and this
results in involves \(n+1\) operations which is still in the order of
\(O(n)\). Functions that require \(n\), \(n+10\) or \(n-2\) operations
are still in the order of \(O(n)\).

Making the loop to obtain the mean worse in terms of efficiency is
useful to clarify the computational complexity. Consider the following
loop in \texttt{R}:

\begin{Shaded}
\begin{Highlighting}[]
\NormalTok{x }\OtherTok{\textless{}{-}} \FunctionTok{c}\NormalTok{(}\DecValTok{1}\NormalTok{, }\DecValTok{2}\NormalTok{, }\DecValTok{3}\NormalTok{, }\DecValTok{4}\NormalTok{, }\DecValTok{5}\NormalTok{)}

\NormalTok{numerator }\OtherTok{\textless{}{-}} \DecValTok{0}
\NormalTok{denominator }\OtherTok{\textless{}{-}} \FunctionTok{length}\NormalTok{(x)}

\ControlFlowTok{for}\NormalTok{ (i }\ControlFlowTok{in} \DecValTok{1}\SpecialCharTok{:}\NormalTok{denominator) \{}
\NormalTok{  y }\OtherTok{\textless{}{-}}\NormalTok{ (y }\SpecialCharTok{+}\NormalTok{ x[i]) }\SpecialCharTok{/}\NormalTok{ denominator}
\NormalTok{\}}
\end{Highlighting}
\end{Shaded}

This loop still has a complexity of \(O(n)\), but it is less efficient
because it involves \(n>1\) divisions, leading to \(kn\) total
operations with \(k>1\). For the big-O notation, \(2n\), \(3n\) or
\(200n + 100\) are also in the order of \(O(n)\).

Regardless of the type of operation, big-O counts the number of
operations. For example, the geometric mean is computationally more
expensive because multiplication (and division) are slower than sums,
but its complexity is still \(O(n)\).

Consider the following loop in C++:

\begin{Shaded}
\begin{Highlighting}[]
\DataTypeTok{int}\NormalTok{ sum }\OperatorTok{=} \DecValTok{0}\OperatorTok{;}
\DataTypeTok{int}\NormalTok{ n }\OperatorTok{=} \DecValTok{9}\OperatorTok{;}
\ControlFlowTok{for} \OperatorTok{(}\DataTypeTok{int}\NormalTok{ i }\OperatorTok{=} \DecValTok{0}\OperatorTok{;}\NormalTok{ i }\OperatorTok{\textless{}}\NormalTok{ n}\OperatorTok{;}\NormalTok{ i}\OperatorTok{++)} \OperatorTok{\{}
\NormalTok{  sum}\OperatorTok{++;}
\OperatorTok{\}}
\end{Highlighting}
\end{Shaded}

The total number of operations is \(3n+1\), because there is one
operation to set the initial value of \texttt{sum}, one to set the
initial value of \texttt{n}, one to set the initial value of \texttt{i},
\(n\) verifications that \(i<n\), \(n\) increments of \texttt{i}, and
\(n\) increases of \texttt{sum} by a value of one. The complexity is
still in the order of \(O(n)\).

Other operations can be more expensive, such as matrix multiplication,
which has a complexity of \(O(n^3)\) for two \(n \times n\) matrices,
and finding the inverse of a matrix, which has a complexity of
\(O(n^3)\) for a \(n \times n\) matrix.

\section{Reduced forms}\label{reduced-forms}

Two function can reach the same result but with a different number of
operations, and therefore different complexity. Adapting from Emara
(2024), here are two function written by Dante and Virgilio to compute
\(2^n\) using recursion:

\begin{Shaded}
\begin{Highlighting}[]
\DataTypeTok{int}\NormalTok{ Dante}\OperatorTok{(}\DataTypeTok{int}\NormalTok{ n}\OperatorTok{)} \OperatorTok{\{}
  \ControlFlowTok{if} \OperatorTok{(}\NormalTok{n }\OperatorTok{==} \DecValTok{0}\OperatorTok{)} \OperatorTok{\{}
    \ControlFlowTok{return} \DecValTok{1}\OperatorTok{;}
  \OperatorTok{\}} \ControlFlowTok{else} \OperatorTok{\{}
    \ControlFlowTok{return} \OperatorTok{(}\NormalTok{Dante}\OperatorTok{(}\NormalTok{n}\OperatorTok{{-}}\DecValTok{1}\OperatorTok{)} \OperatorTok{+}\NormalTok{ Dante}\OperatorTok{(}\NormalTok{n}\OperatorTok{{-}}\DecValTok{1}\OperatorTok{));}
  \OperatorTok{\}}
\OperatorTok{\}}

\DataTypeTok{int}\NormalTok{ Virgilio}\OperatorTok{(}\DataTypeTok{int}\NormalTok{ n}\OperatorTok{)} \OperatorTok{\{}
  \ControlFlowTok{if} \OperatorTok{(}\NormalTok{n }\OperatorTok{==} \DecValTok{0}\OperatorTok{)} \OperatorTok{\{}
    \ControlFlowTok{return} \DecValTok{1}\OperatorTok{;}
  \OperatorTok{\}} \ControlFlowTok{else} \OperatorTok{\{}
    \ControlFlowTok{return} \OperatorTok{(}\DecValTok{2} \OperatorTok{*}\NormalTok{ Virgilio}\OperatorTok{(}\NormalTok{n}\OperatorTok{{-}}\DecValTok{1}\OperatorTok{));}
  \OperatorTok{\}}
\OperatorTok{\}}
\end{Highlighting}
\end{Shaded}

The two functions are correct and equivalent, but the number of
operations is different. \texttt{Dante()} uses two recursive calls for
each step, creating \(2^n\) total calls, and therefore its complexity is
\(O(2^n)\). \texttt{Virgilio()} uses one recursive call for each step,
creating \(n\) total calls, and therefore its complexity is \(O(n)\).

The second function is a reduced form of the first, and in practical
terms it is the same as simplifying \(x^2 + 2x + 1\) to \((x+1)^2\) to
reduce the number of operations required to obtain the result.
\texttt{R} and Armadillo internals make an extensive use of reduced
forms to optimize the code, and in general it is not simple to write a
reduced form in any language, especially for complex operations such as
regression models where the reductions can introduce a wide range of
issues (e.g., unintended divisions by zero in corner cases).

\section{Gauss-Jordan C++
implementation}\label{gauss-jordan-c-implementation}

Consider the following system of linear equations from Vargas Sepúlveda
(2023):

\begin{equation*}
\begin{bmatrix}
1 & 0 & 0 \\
1 & 1 & 0 \\
0 & 1 & 1
\end{bmatrix}
\begin{bmatrix}
x_1 \\
x_2 \\
x_3
\end{bmatrix}
=
\begin{bmatrix}
6.50 \\
7.50 \\
8.50
\end{bmatrix}
\end{equation*}

The system can be solved with row operations to obtain the inverse
matrix: \begin{equation*}
\overset{\text{row 2} - \text{row 1}}{\rightarrow} \begin{bmatrix}
1 & 0 & 0 & | & 1 & 0 & 0 \\
0 & 1 & 0 & | & -1 & 1 & 0 \\
0 & 1 & 1 & | & 0 & 0 & 1
\end{bmatrix} \newline
\overset{\text{row 3} - \text{row 2}}{\rightarrow} \begin{bmatrix}
1 & 0 & 0 & | & 1 & 0 & 0 \\
0 & 1 & 0 & | & -1 & 1 & 0 \\
0 & 0 & 1 & | & 1 & -1 & 1
\end{bmatrix}
\end{equation*}

\begin{equation*}
\begin{bmatrix}
1 & 0 & 0 \\
-1 & 1 & 0 \\
1 & -1 & 1
\end{bmatrix}
\begin{bmatrix}
6.50 \\
7.50 \\
8.50
\end{bmatrix}
=
\begin{bmatrix}
6.50 \\
1.00 \\
7.50
\end{bmatrix}
\end{equation*}

The same can be done with a naive implementation of the Gauss-Jordan
algorithm that has complexity \(O(n^3)\) (Strang 1988). It should serve
as a starting point to understand the syntax and the data structures.

\begin{Shaded}
\begin{Highlighting}[]
\PreprocessorTok{\#include }\ImportTok{\textless{}cpp11.hpp\textgreater{}}

\KeywordTok{using} \KeywordTok{namespace}\NormalTok{ cpp11}\OperatorTok{;}

\OperatorTok{[[}\AttributeTok{cpp11}\OperatorTok{::}\AttributeTok{register}\OperatorTok{]]}\NormalTok{ doubles\_matrix}\OperatorTok{\textless{}\textgreater{}} \VariableTok{invert\_matrix\_}\OperatorTok{(}\NormalTok{doubles\_matrix}\OperatorTok{\textless{}\textgreater{}}\NormalTok{ a}\OperatorTok{)} \OperatorTok{\{}
  \CommentTok{// Check dimensions}
  \DataTypeTok{int}\NormalTok{ n }\OperatorTok{=}\NormalTok{ a}\OperatorTok{.}\NormalTok{nrow}\OperatorTok{(),}\NormalTok{ m }\OperatorTok{=}\NormalTok{ a}\OperatorTok{.}\NormalTok{ncol}\OperatorTok{();}
  \ControlFlowTok{if} \OperatorTok{(}\NormalTok{n }\OperatorTok{!=}\NormalTok{ m}\OperatorTok{)} \OperatorTok{\{}
\NormalTok{    stop}\OperatorTok{(}\StringTok{"X must be a square matrix"}\OperatorTok{);}
  \OperatorTok{\}}

  \CommentTok{// Copy the matrix}
\NormalTok{  writable}\OperatorTok{::}\NormalTok{doubles\_matrix}\OperatorTok{\textless{}\textgreater{}}\NormalTok{ acopy}\OperatorTok{(}\NormalTok{n}\OperatorTok{,}\NormalTok{ n}\OperatorTok{);}
  \ControlFlowTok{for} \OperatorTok{(}\DataTypeTok{int}\NormalTok{ i }\OperatorTok{=} \DecValTok{0}\OperatorTok{;}\NormalTok{ i }\OperatorTok{\textless{}}\NormalTok{ n}\OperatorTok{;}\NormalTok{ i}\OperatorTok{++)} \OperatorTok{\{}
    \ControlFlowTok{for} \OperatorTok{(}\DataTypeTok{int}\NormalTok{ j }\OperatorTok{=} \DecValTok{0}\OperatorTok{;}\NormalTok{ j }\OperatorTok{\textless{}}\NormalTok{ n}\OperatorTok{;}\NormalTok{ j}\OperatorTok{++)} \OperatorTok{\{}
\NormalTok{      acopy}\OperatorTok{(}\NormalTok{i}\OperatorTok{,}\NormalTok{ j}\OperatorTok{)} \OperatorTok{=}\NormalTok{ a}\OperatorTok{(}\NormalTok{i}\OperatorTok{,}\NormalTok{ j}\OperatorTok{);}
    \OperatorTok{\}}
  \OperatorTok{\}}

  \CommentTok{// Create the identity matrix as a starting point for Gauss{-}Jordan}
\NormalTok{  writable}\OperatorTok{::}\NormalTok{doubles\_matrix}\OperatorTok{\textless{}\textgreater{}}\NormalTok{ ainv}\OperatorTok{(}\NormalTok{n}\OperatorTok{,}\NormalTok{ n}\OperatorTok{);}
  \ControlFlowTok{for} \OperatorTok{(}\DataTypeTok{int}\NormalTok{ i }\OperatorTok{=} \DecValTok{0}\OperatorTok{;}\NormalTok{ i }\OperatorTok{\textless{}}\NormalTok{ n}\OperatorTok{;}\NormalTok{ i}\OperatorTok{++)} \OperatorTok{\{}
    \ControlFlowTok{for} \OperatorTok{(}\DataTypeTok{int}\NormalTok{ j }\OperatorTok{=} \DecValTok{0}\OperatorTok{;}\NormalTok{ j }\OperatorTok{\textless{}}\NormalTok{ n}\OperatorTok{;}\NormalTok{ j}\OperatorTok{++)} \OperatorTok{\{}
\NormalTok{      ainv}\OperatorTok{(}\NormalTok{i}\OperatorTok{,}\NormalTok{ j}\OperatorTok{)} \OperatorTok{=} \OperatorTok{(}\NormalTok{i }\OperatorTok{==}\NormalTok{ j}\OperatorTok{)} \OperatorTok{?} \FloatTok{1.0} \OperatorTok{:} \FloatTok{0.0}\OperatorTok{;}
    \OperatorTok{\}}
  \OperatorTok{\}}

  \CommentTok{// Overwrite Ainv by steps with the inverse of A}
  \CommentTok{// (find the echelon form of A)}
  \ControlFlowTok{for} \OperatorTok{(}\DataTypeTok{int}\NormalTok{ i }\OperatorTok{=} \DecValTok{0}\OperatorTok{;}\NormalTok{ i }\OperatorTok{\textless{}}\NormalTok{ m}\OperatorTok{;}\NormalTok{ i}\OperatorTok{++)} \OperatorTok{\{}
    \DataTypeTok{double}\NormalTok{ aij }\OperatorTok{=}\NormalTok{ acopy}\OperatorTok{(}\NormalTok{i}\OperatorTok{,}\NormalTok{ i}\OperatorTok{);}

    \CommentTok{// Divide the row by the diagonal element}
    \ControlFlowTok{for} \OperatorTok{(}\DataTypeTok{int}\NormalTok{ j }\OperatorTok{=} \DecValTok{0}\OperatorTok{;}\NormalTok{ j }\OperatorTok{\textless{}}\NormalTok{ m}\OperatorTok{;}\NormalTok{ j}\OperatorTok{++)} \OperatorTok{\{}
\NormalTok{      acopy}\OperatorTok{(}\NormalTok{i}\OperatorTok{,}\NormalTok{ j}\OperatorTok{)} \OperatorTok{/=}\NormalTok{ aij}\OperatorTok{;}
\NormalTok{      ainv}\OperatorTok{(}\NormalTok{i}\OperatorTok{,}\NormalTok{ j}\OperatorTok{)} \OperatorTok{/=}\NormalTok{ aij}\OperatorTok{;}
    \OperatorTok{\}}

    \CommentTok{// Subtract the row from the other rows}
    \ControlFlowTok{for} \OperatorTok{(}\DataTypeTok{int}\NormalTok{ j }\OperatorTok{=} \DecValTok{0}\OperatorTok{;}\NormalTok{ j }\OperatorTok{\textless{}}\NormalTok{ m}\OperatorTok{;}\NormalTok{ j}\OperatorTok{++)} \OperatorTok{\{}
      \ControlFlowTok{if} \OperatorTok{(}\NormalTok{i }\OperatorTok{!=}\NormalTok{ j}\OperatorTok{)} \OperatorTok{\{}
\NormalTok{        aij }\OperatorTok{=}\NormalTok{ acopy}\OperatorTok{(}\NormalTok{j}\OperatorTok{,}\NormalTok{ i}\OperatorTok{);}
        \ControlFlowTok{for} \OperatorTok{(}\DataTypeTok{int}\NormalTok{ k }\OperatorTok{=} \DecValTok{0}\OperatorTok{;}\NormalTok{ k }\OperatorTok{\textless{}}\NormalTok{ m}\OperatorTok{;}\NormalTok{ k}\OperatorTok{++)} \OperatorTok{\{}
\NormalTok{          acopy}\OperatorTok{(}\NormalTok{j}\OperatorTok{,}\NormalTok{ k}\OperatorTok{)} \OperatorTok{{-}=}\NormalTok{ acopy}\OperatorTok{(}\NormalTok{i}\OperatorTok{,}\NormalTok{ k}\OperatorTok{)} \OperatorTok{*}\NormalTok{ aij}\OperatorTok{;}
\NormalTok{          ainv}\OperatorTok{(}\NormalTok{j}\OperatorTok{,}\NormalTok{ k}\OperatorTok{)} \OperatorTok{{-}=}\NormalTok{ ainv}\OperatorTok{(}\NormalTok{i}\OperatorTok{,}\NormalTok{ k}\OperatorTok{)} \OperatorTok{*}\NormalTok{ aij}\OperatorTok{;}
        \OperatorTok{\}}
      \OperatorTok{\}}
    \OperatorTok{\}}
  \OperatorTok{\}}

  \ControlFlowTok{return}\NormalTok{ ainv}\OperatorTok{;}
\OperatorTok{\}}

\OperatorTok{[[}\AttributeTok{cpp11}\OperatorTok{::}\AttributeTok{register}\OperatorTok{]]}\NormalTok{ doubles\_matrix}\OperatorTok{\textless{}\textgreater{}} \VariableTok{multiply\_inverse\_}\OperatorTok{(}\NormalTok{doubles\_matrix}\OperatorTok{\textless{}\textgreater{}}\NormalTok{ a}\OperatorTok{,}
\NormalTok{                                                       doubles\_matrix}\OperatorTok{\textless{}\textgreater{}}\NormalTok{ b}\OperatorTok{)} \OperatorTok{\{}
  \CommentTok{// Check dimensions}
  \DataTypeTok{int}\NormalTok{ n1 }\OperatorTok{=}\NormalTok{ a}\OperatorTok{.}\NormalTok{nrow}\OperatorTok{(),}\NormalTok{ m1 }\OperatorTok{=}\NormalTok{ a}\OperatorTok{.}\NormalTok{ncol}\OperatorTok{(),}\NormalTok{ n2 }\OperatorTok{=}\NormalTok{ b}\OperatorTok{.}\NormalTok{nrow}\OperatorTok{(),}\NormalTok{ m2 }\OperatorTok{=}\NormalTok{ b}\OperatorTok{.}\NormalTok{ncol}\OperatorTok{();}
  \ControlFlowTok{if} \OperatorTok{(}\NormalTok{n1 }\OperatorTok{!=}\NormalTok{ m1}\OperatorTok{)} \OperatorTok{\{}
\NormalTok{    stop}\OperatorTok{(}\StringTok{"a must be a square matrix"}\OperatorTok{);}
  \OperatorTok{\}}
  \ControlFlowTok{if} \OperatorTok{(}\NormalTok{n1 }\OperatorTok{!=}\NormalTok{ n2}\OperatorTok{)} \OperatorTok{\{}
\NormalTok{    stop}\OperatorTok{(}\StringTok{"b must have the same number of rows as a"}\OperatorTok{);}
  \OperatorTok{\}}
  \ControlFlowTok{if} \OperatorTok{(}\NormalTok{m2 }\OperatorTok{!=} \DecValTok{1}\OperatorTok{)} \OperatorTok{\{}
\NormalTok{    stop}\OperatorTok{(}\StringTok{"b must be a column vector"}\OperatorTok{);}
  \OperatorTok{\}}

  \CommentTok{// Obtain the inverse}
\NormalTok{  doubles\_matrix}\OperatorTok{\textless{}\textgreater{}}\NormalTok{ ainv }\OperatorTok{=} \VariableTok{invert\_matrix\_}\OperatorTok{(}\NormalTok{a}\OperatorTok{);}

  \CommentTok{// Multiply ainv by b}
\NormalTok{  writable}\OperatorTok{::}\NormalTok{doubles\_matrix}\OperatorTok{\textless{}\textgreater{}}\NormalTok{ x}\OperatorTok{(}\NormalTok{n1}\OperatorTok{,} \DecValTok{1}\OperatorTok{);}
  \ControlFlowTok{for} \OperatorTok{(}\DataTypeTok{int}\NormalTok{ i }\OperatorTok{=} \DecValTok{0}\OperatorTok{;}\NormalTok{ i }\OperatorTok{\textless{}}\NormalTok{ n1}\OperatorTok{;}\NormalTok{ i}\OperatorTok{++)} \OperatorTok{\{}
\NormalTok{    x}\OperatorTok{(}\NormalTok{i}\OperatorTok{,} \DecValTok{0}\OperatorTok{)} \OperatorTok{=} \FloatTok{0.0}\OperatorTok{;}
    \ControlFlowTok{for} \OperatorTok{(}\DataTypeTok{int}\NormalTok{ j }\OperatorTok{=} \DecValTok{0}\OperatorTok{;}\NormalTok{ j }\OperatorTok{\textless{}}\NormalTok{ n1}\OperatorTok{;}\NormalTok{ j}\OperatorTok{++)} \OperatorTok{\{}
\NormalTok{      x}\OperatorTok{(}\NormalTok{i}\OperatorTok{,} \DecValTok{0}\OperatorTok{)} \OperatorTok{+=}\NormalTok{ ainv}\OperatorTok{(}\NormalTok{i}\OperatorTok{,}\NormalTok{ j}\OperatorTok{)} \OperatorTok{*}\NormalTok{ b}\OperatorTok{(}\NormalTok{j}\OperatorTok{,} \DecValTok{0}\OperatorTok{);}
    \OperatorTok{\}}
  \OperatorTok{\}}

  \ControlFlowTok{return}\NormalTok{ x}\OperatorTok{;}
\OperatorTok{\}}
\end{Highlighting}
\end{Shaded}

The code above includes the \texttt{cpp11} library
(\texttt{\#include\ \textless{}cpp11.hpp\textgreater{}}) and loads the
corresponding namespace (\texttt{using\ namespace\ cpp11}) to simplify
the notation (e.g., typing
\texttt{doubles\_matrix\textless{}\textgreater{}} instead of
\texttt{cpp11::doubles\_matrix\textless{}\textgreater{}}). It declares
two functions, \texttt{invert\_matrix\_} reads a matrix from \texttt{R}
in a direct way (by making a copy) and returns its inverse and
\texttt{multiply\_inverse\_} that reads from \texttt{R} (\texttt{a} and
\texttt{b}) and C++ (\texttt{ainv}), and solves a system of linear
equations.

These functions use \texttt{doubles\_matrix\textless{}\textgreater{}}
and \texttt{writable::doubles\_matrix\textless{}\textgreater{}}. The
\texttt{writable::} prefix must be added every time the object will be
modified later or the code will not compile. The code was written in a
modular way, organized in two dedicated functions, and resulting object
is assigned to a \texttt{doubles\_matrix\textless{}\textgreater{}} that
goes from C++ to \texttt{R}. In the functions, \texttt{stop()},
\texttt{nrow()} and \texttt{ncol()} are not a part of stock C++, these
are provided by \texttt{cpp11}.

\section{Gauss-Jordan Armadillo
implementation}\label{gauss-jordan-armadillo-implementation}

The previous Gauss-Jordan implementation can be largely simplified by
using the Armadillo library.

\begin{Shaded}
\begin{Highlighting}[]
\PreprocessorTok{\#include }\ImportTok{\textless{}armadillo.hpp\textgreater{}}
\PreprocessorTok{\#include }\ImportTok{\textless{}cpp11.hpp\textgreater{}}
\PreprocessorTok{\#include }\ImportTok{\textless{}cpp11armadillo.hpp\textgreater{}}

\KeywordTok{using} \KeywordTok{namespace}\NormalTok{ arma}\OperatorTok{;}
\KeywordTok{using} \KeywordTok{namespace}\NormalTok{ cpp11}\OperatorTok{;}

\OperatorTok{[[}\AttributeTok{cpp11}\OperatorTok{::}\AttributeTok{register}\OperatorTok{]]}
\NormalTok{doubles\_matrix}\OperatorTok{\textless{}\textgreater{}} \VariableTok{invert\_matrix\_}\OperatorTok{(}\AttributeTok{const}\NormalTok{ doubles\_matrix}\OperatorTok{\textless{}\textgreater{}\&}\NormalTok{ a}\OperatorTok{)} \OperatorTok{\{}
\NormalTok{  Mat}\OperatorTok{\textless{}}\DataTypeTok{double}\OperatorTok{\textgreater{}}\NormalTok{ Acopy }\OperatorTok{=}\NormalTok{ as\_Mat}\OperatorTok{(}\NormalTok{a}\OperatorTok{);}
\NormalTok{  Mat}\OperatorTok{\textless{}}\DataTypeTok{double}\OperatorTok{\textgreater{}}\NormalTok{ Ainv }\OperatorTok{=}\NormalTok{ inv}\OperatorTok{(}\NormalTok{Acopy}\OperatorTok{);}
  \ControlFlowTok{return}\NormalTok{ as\_doubles\_matrix}\OperatorTok{(}\NormalTok{Ainv}\OperatorTok{);}
\OperatorTok{\}}

\OperatorTok{[[}\AttributeTok{cpp11}\OperatorTok{::}\AttributeTok{register}\OperatorTok{]]}
\NormalTok{doubles\_matrix}\OperatorTok{\textless{}\textgreater{}} \VariableTok{multiply\_inverse\_}\OperatorTok{(}\AttributeTok{const}\NormalTok{ doubles\_matrix}\OperatorTok{\textless{}\textgreater{}\&}\NormalTok{ a}\OperatorTok{,}
                                   \AttributeTok{const}\NormalTok{ doubles\_matrix}\OperatorTok{\textless{}\textgreater{}\&}\NormalTok{ b}\OperatorTok{)} \OperatorTok{\{}
\NormalTok{  Mat}\OperatorTok{\textless{}}\DataTypeTok{double}\OperatorTok{\textgreater{}}\NormalTok{ Acopy }\OperatorTok{=}\NormalTok{ as\_Mat}\OperatorTok{(}\NormalTok{A}\OperatorTok{);}
\NormalTok{  Mat}\OperatorTok{\textless{}}\DataTypeTok{double}\OperatorTok{\textgreater{}}\NormalTok{ Bcopy }\OperatorTok{=}\NormalTok{ as\_Mat}\OperatorTok{(}\NormalTok{b}\OperatorTok{);}
\NormalTok{  Mat}\OperatorTok{\textless{}}\DataTypeTok{double}\OperatorTok{\textgreater{}}\NormalTok{ X }\OperatorTok{=}\NormalTok{ inv}\OperatorTok{(}\NormalTok{Acopy}\OperatorTok{)} \OperatorTok{*}\NormalTok{ Bcopy}\OperatorTok{;}
  \ControlFlowTok{return}\NormalTok{ as\_doubles\_matrix}\OperatorTok{(}\NormalTok{X}\OperatorTok{);}
\OperatorTok{\}}
\end{Highlighting}
\end{Shaded}

The \texttt{inv()} function and the \texttt{*} operator verify the
dimensions of the matrices. This example shows that Armadillo largely
simplifies the code. Its enhanced speed is an extra feature to its
readability and conciseness.

\section{Linear models in Armadillo}\label{linear-models-in-armadillo}

One possibility is to start by creating a minimal package with the
provided templates.

\begin{Shaded}
\begin{Highlighting}[]
\FunctionTok{install.packages}\NormalTok{(}\StringTok{"cpp11armadillo"}\NormalTok{)}
\CommentTok{\# or}
\CommentTok{\# remotes::install\_github("pachadotdev/cpp11armadillo")}
\NormalTok{cpp11armadillo}\SpecialCharTok{::}\FunctionTok{create\_package}\NormalTok{(}\StringTok{"armadilloexample"}\NormalTok{)}
\end{Highlighting}
\end{Shaded}

Given a design matrix \(X\) and outcome vector \(y\), one naive function
(available in the package template) to obtain the Ordinary Least Squares
(OLS) estimator \(\hat{\beta} = (X^tX)^{-1}(X^tY)\) (Hansen 2022) as a
matrix (column vector) is:

\begin{Shaded}
\begin{Highlighting}[]
\PreprocessorTok{\#include }\ImportTok{\textless{}armadillo.hpp\textgreater{}}
\PreprocessorTok{\#include }\ImportTok{\textless{}cpp11.hpp\textgreater{}}
\PreprocessorTok{\#include }\ImportTok{\textless{}cpp11armadillo.hpp\textgreater{}}

\KeywordTok{using} \KeywordTok{namespace}\NormalTok{ arma}\OperatorTok{;}
\KeywordTok{using} \KeywordTok{namespace}\NormalTok{ cpp11}\OperatorTok{;}

\OperatorTok{[[}\AttributeTok{cpp11}\OperatorTok{::}\AttributeTok{register}\OperatorTok{]]}
\NormalTok{doubles\_matrix}\OperatorTok{\textless{}\textgreater{}} \VariableTok{ols\_mat\_}\OperatorTok{(}\AttributeTok{const}\NormalTok{ doubles\_matrix}\OperatorTok{\textless{}\textgreater{}\&}\NormalTok{ y}\OperatorTok{,}
                          \AttributeTok{const}\NormalTok{ doubles\_matrix}\OperatorTok{\textless{}\textgreater{}\&}\NormalTok{ x}\OperatorTok{)} \OperatorTok{\{}
\NormalTok{  Mat}\OperatorTok{\textless{}}\DataTypeTok{double}\OperatorTok{\textgreater{}}\NormalTok{ Y }\OperatorTok{=}\NormalTok{ as\_Mat}\OperatorTok{(}\NormalTok{y}\OperatorTok{);}
\NormalTok{  Mat}\OperatorTok{\textless{}}\DataTypeTok{double}\OperatorTok{\textgreater{}}\NormalTok{ X }\OperatorTok{=}\NormalTok{ as\_Mat}\OperatorTok{(}\NormalTok{x}\OperatorTok{);}

  \CommentTok{// \textbackslash{}beta = (X\^{}tX)\^{}\{{-}1\}X\^{}tY}
\NormalTok{  Mat}\OperatorTok{\textless{}}\DataTypeTok{double}\OperatorTok{\textgreater{}}\NormalTok{ b }\OperatorTok{=}\NormalTok{ inv}\OperatorTok{(}\NormalTok{X}\OperatorTok{.}\NormalTok{t}\OperatorTok{()} \OperatorTok{*}\NormalTok{ X}\OperatorTok{)} \OperatorTok{*}\NormalTok{ X}\OperatorTok{.}\NormalTok{t}\OperatorTok{()} \OperatorTok{*}\NormalTok{ Y}\OperatorTok{;}

  \ControlFlowTok{return}\NormalTok{ as\_doubles\_matrix}\OperatorTok{(}\NormalTok{b}\OperatorTok{);}
\OperatorTok{\}}
\end{Highlighting}
\end{Shaded}

The previous code loads the corresponding namespaces (e.g., the
\texttt{using\ namespace\ arma}) in order to simplify the notation
(e.g., using \texttt{Mat} instead of \texttt{arma::Mat}), and then it
declares the function \texttt{ols\_mat()} that takes inputs from
\texttt{R}, does the computation on C++ side, and it can be called from
\texttt{R}. In this particular case, because the output has dimension
\(n\times 1\), it is possible to use \texttt{doubles\ ols\_mat\_} and
\texttt{as\_doubles(beta)} to return an \texttt{R} vector instead of a
matrix.

Unlike the first Gauss-Jordan example, it uses the \texttt{inv()}
function from Armadillo instead of implementing the inverse. It also
uses \texttt{X.t()} to transpose and \texttt{*} to multiply matrices,
which saves writing a loop to transpose and three loops to multiply.
Armadillo uses its own definition of the multiplication operator, and
when it is used with two matrices it does the same as the \texttt{\%*\%}
operator in \texttt{R}.

The use of \texttt{const} and \texttt{\&} are specific to the C++
language and allow to pass data from \texttt{R} to C++ by reference,
that avoid copying the data, and therefore save time and memory.

\texttt{as\_Mat()} and \texttt{as\_doubles\_matrix()} are
\texttt{cpp11armadillo} bridge functions to pass data between \texttt{R}
and the Armadillo library.

In order to use this function in \texttt{R}, it needs to be documented,
and after loading the package it is possible to compare with the
\texttt{R} computation:

\begin{Shaded}
\begin{Highlighting}[]
\NormalTok{devtools}\SpecialCharTok{::}\FunctionTok{document}\NormalTok{()}
\NormalTok{devtools}\SpecialCharTok{::}\FunctionTok{load\_all}\NormalTok{()}

\NormalTok{x }\OtherTok{\textless{}{-}}\NormalTok{ cpp11armadillo}\SpecialCharTok{::}\NormalTok{mtcars\_mat}\SpecialCharTok{$}\NormalTok{x}
\NormalTok{x }\OtherTok{\textless{}{-}}\NormalTok{ x[, }\FunctionTok{c}\NormalTok{(}\StringTok{"wt"}\NormalTok{, }\StringTok{"cyl4"}\NormalTok{, }\StringTok{"cyl6"}\NormalTok{, }\StringTok{"cyl8"}\NormalTok{)]}
\NormalTok{y }\OtherTok{\textless{}{-}}\NormalTok{ cpp11armadillo}\SpecialCharTok{::}\NormalTok{mtcars\_mat}\SpecialCharTok{$}\NormalTok{y}

\FunctionTok{ols\_mat}\NormalTok{(y, x)}
\end{Highlighting}
\end{Shaded}

This can be verified against the \texttt{R} code to verify that the
solution is \(\hat{\beta} = (-3.21, 33.99, 29.74, 27.92)\):

\begin{Shaded}
\begin{Highlighting}[]
\FunctionTok{solve}\NormalTok{(}\FunctionTok{t}\NormalTok{(x) }\SpecialCharTok{\%*\%}\NormalTok{ x) }\SpecialCharTok{\%*\%} \FunctionTok{t}\NormalTok{(x) }\SpecialCharTok{\%*\%}\NormalTok{ y}
\end{Highlighting}
\end{Shaded}

In \texttt{R}, the \texttt{lm()} function does not use a code similar to
the previous implementation. Instead, \texttt{R} uses the QR
decomposition to solve the OLS problem, which is more stable and
efficient than the direct computation of the inverse of \(X^tX\).

A more robust OLS implementation is:

\begin{Shaded}
\begin{Highlighting}[]
\PreprocessorTok{\#include }\ImportTok{\textless{}armadillo.hpp\textgreater{}}
\PreprocessorTok{\#include }\ImportTok{\textless{}cpp11.hpp\textgreater{}}
\PreprocessorTok{\#include }\ImportTok{\textless{}cpp11armadillo.hpp\textgreater{}}

\KeywordTok{using} \KeywordTok{namespace}\NormalTok{ arma}\OperatorTok{;}
\KeywordTok{using} \KeywordTok{namespace}\NormalTok{ cpp11}\OperatorTok{;}

\OperatorTok{[[}\AttributeTok{cpp11}\OperatorTok{::}\AttributeTok{register}\OperatorTok{]]}\NormalTok{ doubles\_matrix}\OperatorTok{\textless{}\textgreater{}} \VariableTok{ols\_mat\_qr\_}\OperatorTok{(}\AttributeTok{const}\NormalTok{ doubles\_matrix}\OperatorTok{\textless{}\textgreater{}\&}\NormalTok{ y}\OperatorTok{,}
                                                 \AttributeTok{const}\NormalTok{ doubles\_matrix}\OperatorTok{\textless{}\textgreater{}\&}\NormalTok{ x}\OperatorTok{)} 
                                                 \OperatorTok{\{}
\NormalTok{  Mat}\OperatorTok{\textless{}}\DataTypeTok{double}\OperatorTok{\textgreater{}}\NormalTok{ Y }\OperatorTok{=}\NormalTok{ as\_Mat}\OperatorTok{(}\NormalTok{y}\OperatorTok{);}
\NormalTok{  Mat}\OperatorTok{\textless{}}\DataTypeTok{double}\OperatorTok{\textgreater{}}\NormalTok{ X }\OperatorTok{=}\NormalTok{ as\_Mat}\OperatorTok{(}\NormalTok{x}\OperatorTok{);}

  \CommentTok{// (X\textquotesingle{}X)\^{}({-}1)}
\NormalTok{  Mat}\OperatorTok{\textless{}}\DataTypeTok{double}\OperatorTok{\textgreater{}}\NormalTok{ Q}\OperatorTok{,}\NormalTok{ R}\OperatorTok{;}
  \DataTypeTok{bool}\NormalTok{ computable }\OperatorTok{=}\NormalTok{ qr\_econ}\OperatorTok{(}\NormalTok{Q}\OperatorTok{,}\NormalTok{ R}\OperatorTok{,}\NormalTok{ X}\OperatorTok{.}\NormalTok{t}\OperatorTok{()} \OperatorTok{*}\NormalTok{ X}\OperatorTok{);}

  \ControlFlowTok{if} \OperatorTok{(!}\NormalTok{computable}\OperatorTok{)} \OperatorTok{\{}
\NormalTok{    stop}\OperatorTok{(}\StringTok{"QR decomposition failed"}\OperatorTok{);}
  \OperatorTok{\}} \ControlFlowTok{else} \OperatorTok{\{}
    \CommentTok{// backsolve \textasciigrave{}R\textasciigrave{}}
\NormalTok{    Mat}\OperatorTok{\textless{}}\DataTypeTok{double}\OperatorTok{\textgreater{}}\NormalTok{ b }\OperatorTok{=}\NormalTok{ solve}\OperatorTok{(}\NormalTok{R}\OperatorTok{,}\NormalTok{ Q}\OperatorTok{.}\NormalTok{t}\OperatorTok{()} \OperatorTok{*}\NormalTok{ X}\OperatorTok{.}\NormalTok{t}\OperatorTok{()} \OperatorTok{*}\NormalTok{ Y}\OperatorTok{);}
    \ControlFlowTok{return}\NormalTok{ as\_doubles\_matrix}\OperatorTok{(}\NormalTok{b}\OperatorTok{);}
  \OperatorTok{\}}
\OperatorTok{\}}
\end{Highlighting}
\end{Shaded}

The previous example, instead of directly inverting \(X^tX\), creates
empty matrices \(Q\) and \(R\), and a boolean (e.g., logical) value for
\texttt{qr\_econ()}, and then the QR function tries to decompose
\(X^tX\) into an orthogonal matrix \(Q\) and an upper triangular matrix
\(R\) such that \(XtX = QR\). If the composition is successful or not,
the function returns ``true'' or ``false'' respectively. The
\texttt{solve()} arguments come from the fact that \(R\) is upper
triangular, and backsolving from the last equation results in
\(R\beta = QX^ty\).

\subsection{Logistic regression}\label{logistic-regression}

Armadillo also provides additional data structures, such as field and
cube, which resemble a list of scalars, vectors or matrices and that are
particularly useful for loops.

Adapting from the OLS examples, it is possible to fit a logistic
regression with a loop that repeats calls to a function that returns the
OLS coefficients. To do this, the starting point is to transform the
data by the logistic link function (McCullagh and Nelder 1989; Vargas
Sepúlveda 2023): \begin{align*}
\mu &= \frac{y + 1/2}{2} \\
\eta &= \log\left(\frac{\mu}{1-\mu}\right) \\
z &= \eta + \frac{y - \mu}{\mu}
\end{align*}

From the transformed outcome \(z\) and a design matrix \(X\), we can
implement a Re-Weighted Least Squares (RWLS) algorithm to obtain the
coefficients of the logistic regression (McCullagh and Nelder 1989). The
following code is a naive implementation of the RWLS algorithm:

\begin{Shaded}
\begin{Highlighting}[]
\PreprocessorTok{\#include }\ImportTok{\textless{}cpp11.hpp\textgreater{}}
\PreprocessorTok{\#include }\ImportTok{\textless{}cpp11armadillo.hpp\textgreater{}}

\KeywordTok{using} \KeywordTok{namespace}\NormalTok{ arma}\OperatorTok{;}
\KeywordTok{using} \KeywordTok{namespace}\NormalTok{ cpp11}\OperatorTok{;}

\NormalTok{Mat}\OperatorTok{\textless{}}\DataTypeTok{double}\OperatorTok{\textgreater{}} \VariableTok{rwls\_mat\_coef\_}\OperatorTok{(}\AttributeTok{const}\NormalTok{ Mat}\OperatorTok{\textless{}}\DataTypeTok{double}\OperatorTok{\textgreater{}\&}\NormalTok{ Y}\OperatorTok{,} \AttributeTok{const}\NormalTok{ Mat}\OperatorTok{\textless{}}\DataTypeTok{double}\OperatorTok{\textgreater{}\&}\NormalTok{ X}\OperatorTok{,}
                          \AttributeTok{const}\NormalTok{ Mat}\OperatorTok{\textless{}}\DataTypeTok{double}\OperatorTok{\textgreater{}\&}\NormalTok{ W}\OperatorTok{)} \OperatorTok{\{}
\NormalTok{  Mat}\OperatorTok{\textless{}}\DataTypeTok{double}\OperatorTok{\textgreater{}}\NormalTok{ Wd }\OperatorTok{=}\NormalTok{ diagmat}\OperatorTok{(}\NormalTok{W}\OperatorTok{);}

  \CommentTok{// \textbackslash{}beta = (X\^{}tWX)\^{}\{{-}1\}X\^{}tWY}
\NormalTok{  Mat}\OperatorTok{\textless{}}\DataTypeTok{double}\OperatorTok{\textgreater{}}\NormalTok{ B }\OperatorTok{=}\NormalTok{ inv}\OperatorTok{(}\NormalTok{X}\OperatorTok{.}\NormalTok{t}\OperatorTok{()} \OperatorTok{*}\NormalTok{ Wd }\OperatorTok{*}\NormalTok{ X}\OperatorTok{)} \OperatorTok{*}\NormalTok{ X}\OperatorTok{.}\NormalTok{t}\OperatorTok{()} \OperatorTok{*}\NormalTok{ Wd }\OperatorTok{*}\NormalTok{ Y}\OperatorTok{;}
  \ControlFlowTok{return}\NormalTok{ B}\OperatorTok{;}
\OperatorTok{\}}

\OperatorTok{[[}\AttributeTok{cpp11}\OperatorTok{::}\AttributeTok{register}\OperatorTok{]]}\NormalTok{ doubles\_matrix}\OperatorTok{\textless{}\textgreater{}} \VariableTok{logistic\_mat\_coef\_}\OperatorTok{(}
    \AttributeTok{const}\NormalTok{ doubles\_matrix}\OperatorTok{\textless{}\textgreater{}\&}\NormalTok{ y}\OperatorTok{,} \AttributeTok{const}\NormalTok{ doubles\_matrix}\OperatorTok{\textless{}\textgreater{}\&}\NormalTok{ x}\OperatorTok{)} \OperatorTok{\{}
  \CommentTok{// v = original variables y and x}
\NormalTok{  field}\OperatorTok{\textless{}}\NormalTok{Mat}\OperatorTok{\textless{}}\DataTypeTok{double}\OperatorTok{\textgreater{}\textgreater{}}\NormalTok{ v }\OperatorTok{=} \OperatorTok{\{}\NormalTok{as\_Mat}\OperatorTok{(}\NormalTok{y}\OperatorTok{),}\NormalTok{ as\_Mat}\OperatorTok{(}\NormalTok{x}\OperatorTok{)\};}

  \CommentTok{// nv = new variables mu, eta, and z}
\NormalTok{  field}\OperatorTok{\textless{}}\NormalTok{Mat}\OperatorTok{\textless{}}\DataTypeTok{double}\OperatorTok{\textgreater{}\textgreater{}}\NormalTok{ nv}\OperatorTok{(}\DecValTok{3}\OperatorTok{);}
\NormalTok{  nv}\OperatorTok{(}\DecValTok{0}\OperatorTok{)} \OperatorTok{=} \OperatorTok{(}\NormalTok{v}\OperatorTok{(}\DecValTok{0}\OperatorTok{)} \OperatorTok{+} \FloatTok{0.5}\OperatorTok{)} \OperatorTok{/} \DecValTok{2}\OperatorTok{;}
\NormalTok{  nv}\OperatorTok{(}\DecValTok{1}\OperatorTok{)} \OperatorTok{=}\NormalTok{ log}\OperatorTok{(}\NormalTok{nv}\OperatorTok{(}\DecValTok{0}\OperatorTok{)} \OperatorTok{/} \OperatorTok{(}\DecValTok{1} \OperatorTok{{-}}\NormalTok{ nv}\OperatorTok{(}\DecValTok{0}\OperatorTok{)));}
\NormalTok{  nv}\OperatorTok{(}\DecValTok{2}\OperatorTok{)} \OperatorTok{=}\NormalTok{ nv}\OperatorTok{(}\DecValTok{1}\OperatorTok{)} \OperatorTok{+} \OperatorTok{(}\NormalTok{v}\OperatorTok{(}\DecValTok{0}\OperatorTok{)} \OperatorTok{{-}}\NormalTok{ nv}\OperatorTok{(}\DecValTok{0}\OperatorTok{))} \OperatorTok{/}\NormalTok{ nv}\OperatorTok{(}\DecValTok{0}\OperatorTok{);}

  \CommentTok{// s = scalars rss1, rss2, dif and tol}
  \CommentTok{// rss = residual sum of squares}
  \CommentTok{// dif = rss1 {-} rss2}
  \CommentTok{// tol = tolerance level for convergence}
  \CommentTok{// initialized with 1, 1, 1, and 0.05 as a starting point}
\NormalTok{  Col}\OperatorTok{\textless{}}\DataTypeTok{double}\OperatorTok{\textgreater{}}\NormalTok{ s }\OperatorTok{=} \OperatorTok{\{}\DecValTok{1}\OperatorTok{,} \DecValTok{1}\OperatorTok{,} \DecValTok{1}\OperatorTok{,} \FloatTok{0.05}\OperatorTok{\};}

  \CommentTok{// res = residuals without and with transformation}
\NormalTok{  field}\OperatorTok{\textless{}}\NormalTok{Mat}\OperatorTok{\textless{}}\DataTypeTok{double}\OperatorTok{\textgreater{}\textgreater{}}\NormalTok{ res}\OperatorTok{(}\DecValTok{2}\OperatorTok{);}

  \CommentTok{// b = regression coefficients}
\NormalTok{  Mat}\OperatorTok{\textless{}}\DataTypeTok{double}\OperatorTok{\textgreater{}}\NormalTok{ b}\OperatorTok{;}

  \ControlFlowTok{while} \OperatorTok{(}\NormalTok{abs}\OperatorTok{(}\NormalTok{s}\OperatorTok{(}\DecValTok{2}\OperatorTok{))} \OperatorTok{\textgreater{}}\NormalTok{ s}\OperatorTok{(}\DecValTok{3}\OperatorTok{))} \OperatorTok{\{}
\NormalTok{    b }\OperatorTok{=} \VariableTok{rwls\_mat\_coef\_}\OperatorTok{(}\NormalTok{nv}\OperatorTok{(}\DecValTok{2}\OperatorTok{),}\NormalTok{ v}\OperatorTok{(}\DecValTok{1}\OperatorTok{),}\NormalTok{ nv}\OperatorTok{(}\DecValTok{0}\OperatorTok{));}
\NormalTok{    res}\OperatorTok{(}\DecValTok{0}\OperatorTok{)} \OperatorTok{=}\NormalTok{ nv}\OperatorTok{(}\DecValTok{2}\OperatorTok{)} \OperatorTok{{-}}\NormalTok{ v}\OperatorTok{(}\DecValTok{1}\OperatorTok{)} \OperatorTok{*}\NormalTok{ b}\OperatorTok{;}
\NormalTok{    nv}\OperatorTok{(}\DecValTok{1}\OperatorTok{)} \OperatorTok{=}\NormalTok{ nv}\OperatorTok{(}\DecValTok{2}\OperatorTok{)} \OperatorTok{{-}}\NormalTok{ res}\OperatorTok{(}\DecValTok{0}\OperatorTok{);}
\NormalTok{    nv}\OperatorTok{(}\DecValTok{0}\OperatorTok{)} \OperatorTok{=}\NormalTok{ exp}\OperatorTok{(}\NormalTok{nv}\OperatorTok{(}\DecValTok{1}\OperatorTok{))} \OperatorTok{/} \OperatorTok{(}\DecValTok{1} \OperatorTok{+}\NormalTok{ exp}\OperatorTok{(}\NormalTok{nv}\OperatorTok{(}\DecValTok{1}\OperatorTok{)));}
\NormalTok{    nv}\OperatorTok{(}\DecValTok{2}\OperatorTok{)} \OperatorTok{=}\NormalTok{ nv}\OperatorTok{(}\DecValTok{1}\OperatorTok{)} \OperatorTok{+} \OperatorTok{(}\NormalTok{v}\OperatorTok{(}\DecValTok{0}\OperatorTok{)} \OperatorTok{{-}}\NormalTok{ nv}\OperatorTok{(}\DecValTok{0}\OperatorTok{))} \OperatorTok{/}\NormalTok{ nv}\OperatorTok{(}\DecValTok{0}\OperatorTok{);}
\NormalTok{    res}\OperatorTok{(}\DecValTok{1}\OperatorTok{)} \OperatorTok{=}\NormalTok{ v}\OperatorTok{(}\DecValTok{0}\OperatorTok{)} \OperatorTok{{-}}\NormalTok{ v}\OperatorTok{(}\DecValTok{1}\OperatorTok{)} \OperatorTok{*}\NormalTok{ b}\OperatorTok{;}
\NormalTok{    s}\OperatorTok{(}\DecValTok{1}\OperatorTok{)} \OperatorTok{=}\NormalTok{ accu}\OperatorTok{(}\NormalTok{res}\OperatorTok{(}\DecValTok{1}\OperatorTok{)} \OperatorTok{\%}\NormalTok{ res}\OperatorTok{(}\DecValTok{1}\OperatorTok{));}
\NormalTok{    s}\OperatorTok{(}\DecValTok{2}\OperatorTok{)} \OperatorTok{=}\NormalTok{ s}\OperatorTok{(}\DecValTok{1}\OperatorTok{)} \OperatorTok{{-}}\NormalTok{ s}\OperatorTok{(}\DecValTok{0}\OperatorTok{);}
\NormalTok{    s}\OperatorTok{(}\DecValTok{1}\OperatorTok{)} \OperatorTok{=}\NormalTok{ s}\OperatorTok{(}\DecValTok{0}\OperatorTok{);}
  \OperatorTok{\}}

\NormalTok{  b }\OperatorTok{=} \VariableTok{rwls\_mat\_coef\_}\OperatorTok{(}\NormalTok{nv}\OperatorTok{(}\DecValTok{2}\OperatorTok{),}\NormalTok{ v}\OperatorTok{(}\DecValTok{1}\OperatorTok{),}\NormalTok{ nv}\OperatorTok{(}\DecValTok{0}\OperatorTok{));}
  \ControlFlowTok{return}\NormalTok{ as\_doubles\_matrix}\OperatorTok{(}\NormalTok{b}\OperatorTok{);}
\OperatorTok{\}}
\end{Highlighting}
\end{Shaded}

\section{Conclusion}\label{conclusion}

\texttt{cpp11armadillo} provides a simple and efficient way to integrate
C++ code with \texttt{R}, leveraging the \texttt{cpp11} package and the
Armadillo library. It simplifies the process of writing C++ code for
\texttt{R} users, allowing them to focus on the logic of the algorithm
rather than the technical details of the integration. It can help to
solve performance bottlenecks in \texttt{R} code by using the efficient
linear algebra operations provided by Armadillo in cases where
vectorization is challenging.

\section{Acknowledgements}\label{acknowledgements}

We would like to thank Professor Salma Emara who taught us C++ in the
course ECE244 (Programming Fundamentals). \texttt{cpp11armadillo} is a
byproduct of the knowledge we acquired in that course.

\section*{References}\label{references}
\addcontentsline{toc}{section}{References}

\phantomsection\label{refs}


\end{document}